\documentclass[aps,prd,showpacs,amssymb,twocolumn,nofootinbib]{revtex4}
\usepackage{graphicx}
\begin{document}

\title{Length Uncertainty in a Gravity's Rainbow Formalism}

\author{Pablo \surname{Gal\'an}}
\affiliation{Instituto de Estructura de la Materia, CSIC, Serrano
121, 28006 Madrid, Spain}

\author{Guillermo A. \surname{Mena Marug\'an}}
\affiliation{Instituto de Estructura de la Materia, CSIC, Serrano
121, 28006 Madrid, Spain}


\begin{abstract}
It is commonly accepted that the combination of quantum mechanics
and general relativity gives rise to the emergence of a minimum
uncertainty both in space and time. The arguments that support this
conclusion are mainly based on perturbative approaches to the
quantization, in which the gravitational interactions of the matter
content are described as corrections to a classical background. In a
recent paper, we analyzed the existence of a minimum time
uncertainty in the framework of doubly special relativity. In this
framework, the standard definition of the energy-momentum of
particles is modified appealing to possible quantum gravitational
effects, which are not necessarily perturbative. Demanding that this
modification be completed into a canonical transformation determines
the implementation of doubly special relativity in position space
and leads to spacetime coordinates that depend on the
energy-momentum of the particle. In the present work, we extend our
analysis to the quantum length uncertainty. We show that, in generic
cases, there actually exists a limit in the spatial resolution, both
when the quantum evolution is described in terms of the auxiliary
time corresponding to the Minkowski background or in terms of the
physical time. These two kinds of evolutions can be understood as
corresponding to perturbative and non-perturbative descriptions,
respectively. This result contrasts with that found for the time
uncertainty, which can be made to vanish in all models with
unbounded physical energy if one adheres to a non-perturbative
quantization.

\end{abstract}


\maketitle
\renewcommand{\theequation}{\arabic{section}.\arabic{equation}}

\section{Introduction}

A standard result in Quantum Mechanics is that the measurement of
the position of a quantum state is affected by an uncertainty that
satisfies the Heisenberg relations \cite{GP}. In order to diminish
the position uncertainty one is thus forced to consider states with
increasing momentum uncertainty, achieving an infinite spatial
resolution only at the cost of completely delocalizing the momentum.
In the presence of gravity, however, the situation becomes more
complicated. Via Einstein equations, an uncertainty in the
(energy-)momentum of the system results in one in the geometry,
which implies an additional uncertainty in the position. The total
position uncertainty will therefore consist in the combined effect
of a purely quantum mechanical contribution and a contribution of
gravitational origin \cite{Luis}. In these circumstances, one should
not expect that an infinite spatial resolution can be reached,
unless there exists a very specific relation between these types of
contributions. Similar conclusions apply to the measurements of
length of spatial intervals, determined by the positions of their
endpoints.

The most common approach to analyze the emergence of a minimum
spatial (or time) uncertainty when gravity comes into the scene
consists in adopting a perturbative scheme. The starting point is a
flat background where the matter is inserted. This matter curves the
spacetime, producing a deformation of the geometry which in turn
modifies the expression of the physical energy and momentum of the
system (usually defined in terms of normalized -asymptotic- Killing
vectors). The process continues with successive corrections that one
assumes to be less and less important. The studies in the literature
indicate that a minimum uncertainty is ineluctable in this kind of
perturbative quantization (at least in the next-to-leading-order
approximation) \cite{Luis,Pad,unc,unc2}. A different issue, which is
still open to debate, is whether the same result holds as well in
the context of a non-perturbative quantum description \cite{PG,BMV}.

A suitable arena to test some of these issues is provided by doubly
special relativity (DSR) \cite{Amelino,Amelino2}. In this kind of
theories, the definition of the physical energy and momentum of
particles is modified with respect to the standard relativistic one
in order to encode, at least to some extent, the possible effects of
the gravitational interactions, without necessarily adhering to any
perturbative interpretation. The modification is such that the
system presents an energy and/or momentum scale which is invariant
under Lorentz transformations. This is possible because the action
of the Lorentz group becomes nonlinear on the physical
energy-momentum space \cite{Amelino,Amelino2,DSR1,DSR2,DSR12}.

Several proposals have been put forward for the realization of DSR
in position space \cite{kappa,position,mignemi}. In a previous paper
\cite{PG} we suggested that this realization should be determined by
completing into a canonical transformation the nonlinear mapping
that relates the original energy-momentum variables of standard
relativity in Minkowski spacetime (that we will call pseudo
variables from now on) with the physical energy-momentum of the
system in DSR \cite{judes}. In this framework, the background
Minkowski coordinates are mapped to a new set of spacetime
coordinates that can be regarded as canonically conjugate to the
physical energy-momentum. Those coordinates are linear in the
Minkowski ones, but depend in a non-trivial way on the energy and
momentum of the particle. Owing to this dependence of the spacetime
description, the formalism can be considered a kind of gravity's
rainbow \cite{rainbow}.

Our discussion in Ref. \cite{PG} was focused on the existence of a
minimum time uncertainty in quantum theories derived from DSR. In
particular, we considered the different possibilities of describing
the quantum evolution in terms of a parameter that corresponds
either to the original time of the Minkowski background or to the
physical time of the system. According to our comments above, we
will respectively refer to these two types of quantization as
perturbative and non-perturbative ones, given the distinct
philosophy in the use of background structures. Our analysis proved
that, while there always exists a non-vanishing uncertainty in the
physical time when a perturbative quantization is adopted, an
infinite time resolution can be achieved in certain theories when
the quantization is non-perturbative. More precisely, no minimum
time uncertainty arises non-perturbatively in DSR theories whose
physical energy is unbounded from above. The aim of the present work
is to extend this study of the uncertainty from time lapses to the
case of spatial intervals.

A particular class of spacetimes in which the commented analysis of
the time uncertainty has been carried out in detail is that of the
Einstein-Rosen waves \cite{BMV}. These linearly polarized waves are
described by cylindrically symmetric spacetimes in 3+1 dimensions,
but can equivalently be described in terms of a massless scalar
field coupled to gravity in 2+1 dimensions with axial symmetry
\cite{ERscalar,ERash,ERBMV}. In this dimensionally reduced
formulation, the system can in fact be viewed as an example of DSR
theories, with a physical energy that is bounded from above
\cite{2+1,ERbound}. Therefore, for Einstein-Rosen waves, a
non-vanishing quantum time uncertainty emerges both in the
perturbative and in the non-perturbative approaches. The study of
the spatial uncertainty is not specially interesting in this case,
because the associated DSR theory involves no modification in the
definition of the momenta nor in the canonically conjugate position
variables.

The rest of the paper is organized as follows. In the following
section, we review some aspects of the formulation of DSR theories
in momentum space and introduce our canonical proposal for their
realization in position space. We obtain spacetime coordinates that
are conjugate to the physical energy-momentum, arriving at a
gravity's rainbow formalism. Next, we study the quantization of this
formalism, restricting our considerations to free systems that can
be described within a Hamiltonian scheme. Adopting a perturbative
approach to the quantization, we analyze in Sec. III the length
uncertainty, i.e. the uncertainty in the difference of spatial
positions. We show in Sec. IV that this uncertainty cannot vanish in
the perturbative case under quite generic assumptions. Furthermore,
in Sec. V we prove that the appearance of a minimum length
uncertainty persists when the quantum evolution is described in
terms of the physical time, i.e., in a non-perturbative
quantization. However, we comment the possibility that in some DSR
models one could construct a different type of non-perturbative
quantum theory where the physical position operator became
explicitly time independent. In this scenario, the resolution in the
spatial position could in principle be made as large as desired if
the DSR theory does not possess an invariant momentum scale. The
uncertainty in the physical length (as well as in the physical time
lapse) is studied in Sec. VI in the low-energy sector, approximating
the results of the perturbative quantization up to first order
corrections. In Sec. VII we consider the \emph{massless} case in
this approximation for large values of the Minkowski time $T$. We
show that the uncertainty increases then like the square root of
$T$, just as it occurs in Salecker and Wigner devices \cite{SalWig}.
We present our conclusions in Sec. VIII. Finally, two appendices are
added. In the following, we will adopt units in which $\hbar=c=1$
(with $\hbar$ being Planck constant and $c$ the speed of light).

\section{DSR in momentum and position spaces}
\setcounter{equation}{0}

A characteristic feature of DSR theories is that they possess a
Lorentz invariant energy and/or momentum scale, apart from the scale
provided in standard relativity by the speed of light
\cite{Amelino,Amelino2,DSR1,DSR2,DSR12}. The invariance of such a
scale is possible only thanks to a nonlinear realization of the
Lorentz group in momentum space. A simple way to construct a
realization of this kind is by introducing an invertible map $U$
between the physical energy-momentum $P^{a}=(E,p^i)$ and a standard
Lorentz 4-vector $\Pi^{a}=(\epsilon,\pi^i)$, which we call the
pseudo energy-momentum \cite{judes} (lowercase Latin indices from
the beginning and the middle of the alphabet represent Lorentz and
flat spatial indices, respectively). Denoting the usual linear
action of the Lorentz group by $\mathcal{L}$, the nonlinear Lorentz
transformations are then given by $L(P) =
(U^{-1}\circ\mathcal{L}\circ U)(P)$ \cite{judes,MS}.

The map $U$ must reduce to the identity when energies and momenta
are negligibly small compared to the DSR scale, so that the physical
and pseudo variables coincide in this limit. In addition, a
simplifying assumption that is generally accepted is that the
standard action of rotations is preserved; only boosts are modified
in DSR \cite{MS,kappa}. So, with the notation $p:=|\vec{p}\,|$ and
$\pi:=|\vec{\pi}|$, the most general expression for the map $U$
becomes \cite{kappa,PG}
\begin{eqnarray}\label{momenta}
\Pi = U(P) &\Rightarrow& \left\{ \begin{array}{l} \epsilon=
\tilde{g}(E,p)\,,\\
\pi^i = \tilde{f}(E,p)\,\frac{p^i}{p}\,,
\end{array} \right. \nonumber\\
P = U^{-1}(\Pi) &\Rightarrow& \left\{ \begin{array}{l} E =
g(\epsilon,\pi)\,,\\
p^i =f(\epsilon,\pi)\,\frac{\pi^i}{\pi}\,.
\end{array} \right.
\end{eqnarray}

Since the only invariant energy-momentum scale in standard special
relativity is at infinity, the DSR theory admits a Lorentz invariant
scale at a finite value of the energy and/or momentum only if the
map $U$ has a singularity there \cite{MS}. The domain of definition
of $U$ (which is assumed to contain the low energy-momentum sector)
is therefore bounded from above by that scale. Consequently, DSR
theories can be classified in three types: DSR1 if it is only the
physical momentum that is bounded from above, DSR3 if it is the
physical energy what is bounded, and DSR2 if both the physical
energy and momentum are bounded.

As it is implicit in our discussion, DSR theories are usually
formulated in momentum space, mainly owing to the increasing
interest in investigating the observational implications of deformed
dispersion relations \cite{Amelino,phenomen}. There are different
proposals to determine what is the modified spacetime geometry and
the corresponding transformation rules in position space that should
complement this formulation \cite{kappa,position}. Among them, one
of the most popular consists in abandoning the commutativity of the
spacetime coordinates, as it happens e.g. in $\kappa$-deformed
Minkowski spacetime \cite{DSR12,kappa}.

However, noncommutative geometries are by no means the only way to
obtain a consistent realization in position space. The same goal can
be achieved without renouncing the conventional framework of
commutative spacetimes. In fact, the literature contains several
suggestions for realizations of this kind
\cite{position,mignemi,PG,hinterleitner}. A particular example was
put forward by Magueijo and Smolin \cite{rainbow}, who required that
the contraction between the energy-momentum and an infinitesimal
spacetime displacement were a linear invariant in DSR. This
requirement leads to new spacetime coordinates that depend on the
energy-momentum. Ultimately, the system adopts a spacetime metric
that directly depends on the energy and momentum of its particle
content. This explains the name of \emph{gravity's rainbow} that has
been given to this class of DSR implementations.

In this work, we will follow a suggestion for the realization of DSR
in position space that differs from that of Magueijo and Smolin,
although it leads as well to a gravity's rainbow formalism in the
sense of the energy dependence of the geometry. We will adopt the
proposal of Ref. \cite{PG}, namely, we will specify the realization
by demanding the invariance of the symplectic form ${\bf d}q^a
\wedge {\bf d} \Pi_a$ (where the wedge denotes the exterior product
and Lorentz indices are lowered with the Minkowski metric). This
assigns to the system new, modified spacetime coordinates $x^a$ that
are conjugate to the physical energy-momentum $P_a$, so that the
relation between $(q^a,\Pi_a)$ and $(x^a,P_a)$ is given by a
canonical transformation. Similar proposals for a canonical
implementation of DSR theories have been analyzed by other authors
\cite{mignemi,hinterleitner}.

By completing the map $U$ into a canonical transformation, one
easily derives the following expressions for the new spacetime
coordinates \cite{PG}:
\begin{eqnarray}\label{x}x^i\!\!&=&\!\!\frac{1}{J}\!\left[\partial_{\pi}g\;
\frac{\pi^i}{\pi}q^0\!+\partial_{\epsilon}g\; \frac{\pi^i
\pi_j}{\pi^2}q^j\right]\!+\frac{\pi}
{f}\!\left(q^i\!-\frac{\pi^i\pi_j}{\pi^2}q^j\right)\!, \nonumber\\
x^0\!&=&\!\frac{1}{J}\left[\partial_{\pi}f\;
q^0+\partial_{\epsilon}f\;  \frac{\pi_i}{\pi}q^i\right].
\end{eqnarray}
Here, $J=\partial_{\epsilon} g\partial_{\pi} f -\partial_{\pi}
g\partial_{\epsilon} f$ is the determinant of the Jacobian of the
transformation $U^{-1}$ between $(\epsilon,\pi)$ and $(E,P)$, and
the functions $f$ and $g$ (and therefore $J$) depend on
$(\epsilon,\pi)$. We point out that the transformation (\ref{x}) is
linear in the coordinates $q^a$, but generally depends non-trivially
on the energy-momentum.

We will refer to $(x^a,P_a)$ and $(q^a,\Pi_a)$ as physical and
background (or pseudo) variables, respectively, and will denote
$q^0$ by $T$ and $x^0$ by $t$ to emphasize the role played by the
evolution parameter in our discussion. In addition, we assume in the
following that the system admits a Hamiltonian description, so that
the value of the physical and pseudo energies are respectively given
by a physical Hamiltonian $H$ and a background Hamiltonian $H_0$.
Together with Eq. (\ref{momenta}), we then get $E \rightarrow
H=g(H_0,\pi)$ and $\epsilon\rightarrow H_0=\tilde{g}(H,p)$. Finally,
since DSR theories are essentially conceived as effective
descriptions of free particles that incorporate quantum
gravitational phenomena, we will concentrate our analysis on free
systems. For such systems, the energy and momentum are constants of
motion. The Hamiltonian is hence time independent and commutes with
the momentum under Poisson brackets, both for the physical and the
background variables.

\section{Physical Length Uncertainty: Perturbative Case}
\setcounter{equation}{0}

In this section, we will consider the perturbative approach to the
quantization of the system in which one adopts the background time
coordinate $q^0=T$ as evolution parameter, so that the evolution is
generated by the Hamiltonian $H_0$. We assume that a quantization of
this kind is feasible. In such a quantum description, the physical
time is represented by a genuine operator $\hat{t}$ \cite{BMV,PG}.
We want to study whether the spatial position and length determined
by the physical coordinates $x^i$ is affected in this case by a
non-vanishing quantum uncertainty. In order to simplify the analysis
and deal only with scalar quantities (circumventing the kind of
problems derived from the use of vector components and their
dependence on choices of fixed background structures, choices which
are questionable both from the viewpoint of general relativity and
of the fluctuations inherent to quantum mechanics) we will focus our
attention exclusively on the projection of the position vector along
the direction of motion:
\begin{equation}\label{projection} X:=x^i \frac{p_i}{p}=
x^i\frac{\pi_i}{\pi}=\frac{1}{J}\left[\partial_{\pi} g\;T +
\partial_{H_0} g\,\frac{\pi_j}{\pi}q^j \right].
\end{equation} We recall that $g$, $f$, and $J$ are functions of
only $H_0$ and $\pi$. Remarkably, this expression is similar to that
given in (\ref{x}) for the time coordinate $x^0=t$ with the exchange
of the function $f$ for $g$ and a flip of global sign (so that the
determinant of the Jacobian $J$ is preserved under the commented
exchange).

Given our restriction to free systems, where the energy and momentum
are conserved, the only variable in the expression for $X$ that
evolves in time (apart from the parameter $T$) is
\begin{equation}\label{sT} s_T:=\pi_jq^j.\end{equation}
The subscript $T$ emphasizes this time dependence. Moreover, since
the system is free, the background Hamiltonian $H_0$ is a function
of only the pseudo momentum. Then, from the Hamiltonian equations of
motion, the time derivative of $s_T$ equals $\pi H_0^{\prime}$,
which is a constant of motion. Here, the prime denotes the
derivative with respect to $\pi$. Thus, we conclude that $s_T=s_0+T
\pi H_0^{\prime}$, where $s_0$ is the value of $s_T$ at the initial
instant of time.

For our quantum analysis we will only consider differences between
position variables, avoiding in this way the arbitrariness in the
choice of an origin and the conceptual tensions that arise from
fixing it classically while allowing quantum fluctuations in the
spatial position. The physics of the problem suggests two possible
elections of reference for the position, namely, either the physical
or the background initial value (of the projection along the
direction of motion) of the position vector. In the first case, the
position difference determines the physical interval covered by the
particle in the background lapse $T$. In the second case the
difference includes as well the effective corrections to the initial
background position contained in DSR. We will study both
possibilities to show that our conclusions do not depend on the
specific choice adopted. To distinguish between the two cases, we
introduce a parameter $\eta$, with $\eta=0$ corresponding to the
initial physical position and $\eta=1$ to the background one.
Explicitly, the former of these positions is given by Eq.
(\ref{projection}) with $T=0$ and $\pi_jq^j$ replaced with $s_0$,
whereas the latter is equal to $s_0/\pi$.

From the difference between $X$ and any of these reference
positions, we obtain the following length:
\begin{eqnarray} L_{\eta}&:=&
\frac{1}{J}\left[\partial_{\pi} g\;T + \partial_{H_0} g\,
S_{T}+\eta\,\frac{(\partial_{H_0}g-J)}{\pi} s_0\right],
\nonumber\\
\label{length} S_{T}&:=&\frac{s_T-s_0}{\pi}.
\end{eqnarray}
We will refer to it as the physical length. To represent it as an
operator, we write
\begin{eqnarray}
\label{X}\widehat{L}_{\eta}&:=&\widehat{M}(H_0,\pi)\,T+
\widehat{R}_{T,\eta},\\ \nonumber\\
\label{C}\widehat{R}_{T,\eta}&=&\frac{1}{2}\left(\widehat{N}(H_0,\pi)
\widehat{S}_{T}
+\widehat{S}_{T}\widehat{N}(H_0,\pi)\right)\nonumber\\
&+&\frac{\eta}{2}\left(\widehat{O}(H_0,\pi)\widehat{s}_0+
\widehat{s}_0\widehat{O}(H_0,\pi)\right),
\end{eqnarray}
where
\begin{equation}
\label{A} M:=\frac{\partial_{\pi} g}{J}\;,\quad
N:=\frac{\partial_{H_0} g}{J},\quad O:=\frac{\partial_{H_0}g-J}{\pi
J}.
\end{equation}
The subscript $T$ denotes again dependence on time. In Eqs.
(\ref{X}) and (\ref{C}), we have symmetrized the products of
$\widehat{N}$ with $\widehat{S}_{T}$ and $\widehat{O}$ with
$\widehat{s}_0$, and displayed explicitly the arguments of the
functions $M$, $N$, and $O$. As we have commented, these functions
correspond to constants of motion. Their respective operators can be
defined in terms of those for $H_0$ and $\pi$ employing the spectral
theorem. As for the operator representing $s_{T}$ (and hence
$S_{T}$), we will comment on its definition later in this section.

It is worth pointing out that our expressions are to some extent
similar to those introduced in Ref. \cite{PG} for the physical time
operator $\hat{t}$. The differences come from the fact that in the
latter case the role of the initial background position variable
$s_0/\pi$ is played by the initial background time ($T=0$), and that
in that work we only analyzed the choice $\eta=1$ (initial time
identified with that of the background time parameter). Our analysis
here can be easily applied to the resulting time lapse, $t_{\eta}$,
the precise correspondence being the disappearance of the
contribution $-1/\pi$ in the function $O(H_0,\pi)$ (and therefore in
$ \widehat{R}_{T,\eta}$), the exchange of the function $f$ for $g$
in the resulting formulas, and a flip of global sign.

In order to calculate the uncertainty in the physical length
operator $\widehat{L}_{\eta}$, we will follow the same procedure
that was explained in Ref. \cite{PG}. Given a quantum state, one can
measure the probability densities of any set of observables at any
instant of time \cite{note0}. In this way, one can determine e.g.
the expectation value of those operators. In addition, one can
estimate the value of the parameter $T$ at that instant of time by
analyzing the evolution of the probability densities of observables
in the considered state. This procedure allows to derive a
statistical distribution for $T$ with probability density $\rho(T)$
(and mean value $\bar{T}$). Heisenberg relations imply that the
uncertainty $\Delta T$ of this distribution satisfies the inequality
$\Delta T\Delta H_0\geq 1/2$ (usually called the fourth Heisenberg
relation) \cite{GP,PG}. The double average process involved by the
quantum expectation value $\langle\;\rangle$ and by the estimation
of the time parameter leads to the following uncertainty:
\begin{equation} \label{uncert}
(\Delta L_{\eta})^2\!=\!\!\int dT \rho (T) \big\langle \left(
\widehat{M}\,T+ \widehat{R}_{T,\eta}- \langle \widehat{M} \rangle
\overline{T}- \langle \widehat{R}_{\overline{T},\eta} \rangle
\right)^2 \!\big\rangle.
\end{equation}
Here, $\langle\widehat{R}_{\overline{T},\eta}\rangle$ is the mean
value of the operator $\widehat{R}_{T,\eta}$ computed with the
commented double average \cite{PG}.

At this stage, some remarks are in order about the precise operator
representation adopted for $s_{T}$ when defining
$\widehat{R}_{T,\eta}$ and how this affects the measurements that
are necessary to determine the mean value of this observable. Two
cases are worth commenting. On the one hand, one can represent $s_T$
as an explicitly $T$-independent operator by simply adopting a
symmetrized factor ordering in Eq. (\ref{sT}) and directly promoting
the canonical background variables $(q^i,\pi_i)$ to operators.
Similarly, we can define $\widehat{S}_{T}$ from its symmetrized
classical expression. By performing quantum measurements at the
fixed instant of time in which the system is analyzed, one can then
determine the probability distribution for $s_T$ at that instant. No
estimation of the value of the evolution parameter is needed, so
that the average over $T$ becomes spurious. Similar arguments apply
to the products of $s_{T}$ with constants of motion that appear in
$\widehat{R}_{T,\eta}$. At least in principle, one may hence
identify $\langle\widehat{R}_{\overline{T},\eta}\rangle$ and
$\langle\widehat{R}_{T,\eta}\rangle$ in Eq. (\ref{uncert}), even if
the exact value of $T$ in which the measurements are made is not
known.

On the other hand, one can instead reflect explicitly all the
$T$-dependence of $s_T$ in the definition of its associated
operator. Starting with the solution to its evolution equation, one
arrives at
$\widehat{s}_T:=\widehat{s}_0+T\,\widehat{\pi}\widehat{H_0^{\prime}}$.
So $\widehat{S}_T:=T\,\widehat{H_0^{\prime}}$. Here,
$\widehat{H_0^{\prime}}$ can be defined in terms of the pseudo
momentum using the spectral theorem. Since the operator
$\widehat{H_0^{\prime}}$ corresponds to a constant of motion, its
probability density does not evolve in time. Actually, the same
happens with $\widehat{s}_0$, $\widehat{M}$, $\widehat{N}$ and
$\widehat{O}$, appearing in Eqs. (\ref{X}) and (\ref{C}). In
particular, the measurements of all of their densities can be
performed at an initial instant of time, identified with $T=0$. For
all other instants, the only missing piece of information is the
probability density $\rho(T)$, obtained through measurements of
distributions of observables that track the passage of time. In this
case, obviously, the average with $\rho(T)$ cannot be obviated when
calculating the mean value of $\widehat{R}_{T,\eta}$.

The two cases can nevertheless be studied in exactly the same way by
simply combining all the explicit linear $T$-dependence of
$\widehat{X}$. In the latter case, one gets
\begin{eqnarray}
\label{X2}\widehat{L}_{\eta}\!&=&\!\widehat{\,Y}(H_0,\pi)\,T+
\widehat{Z}_{\eta}(H_0,\pi,s_0),\\
\label{V}\widehat{\,Y}(H_0,\pi)\!&=&\!\widehat{M}(H_0,\pi)+
\widehat{N}(H_0,\pi)\,\widehat{H_0^{\prime}}(\pi),\\
\label{W}\hspace*{-.5cm}\widehat{Z}_{\eta}(H_0,\pi,s_0)\!&=&\!\frac{\eta}{2}
\left(\widehat{O}(H_0,\pi)\widehat{s}_0
+\widehat{s}_0\widehat{O}(H_0,\pi)\right).
\end{eqnarray}
For computational purposes, expression (\ref{X}) can be considered a
particular example of formula Eq. (\ref{X2}) with
$\widehat{\,Y}=\widehat{M}$ and
$\widehat{Z}_{\eta}=\widehat{R}_{T,\eta}$. With the same
substitutions in Eq (\ref{uncert}), the physical length uncertainty
can then be rewritten:
\begin{equation}\label{unc}
(\Delta L_{\eta})^2=[\Delta(Y\overline{T}+Z_{\eta})]^2 +\langle
\widehat{\,Y}\rangle^2(\Delta T)^2+(\Delta T\Delta Y)^2.
\end{equation}
The case of the physical time lapse can be treated in a completely
similar way \cite{PG}, removing the contribution $-1/\pi$ to $O$ in
the definition of $Z_{\eta}$, interchanging the functions $f$ and
$g$, and introducing a global change of sign (to preserve that of
$J$).

\section{Existence of a Minimum Uncertainty in the Perturbative Case}
\setcounter{equation}{0}

The physical length uncertainty vanishes if and only if the three
positive terms that form the r.h.s. of equation (\ref{unc}) are
equal to zero. We will show in this section that this cannot
generally occur.

In order for the uncertainty to vanish, it must in particular do so
at large $T$, times for which the contribution $(\overline{T}\Delta
Y)^2$ dominates in (\ref{unc}). Therefore, $\Delta Y$ (which is
independent of time) must vanish. Let us assume that the expression
of the background Hamiltonian $H_0$ as a function of $\pi$ is
invertible for the whole range of pseudo energies, i.e. $\pi=\pi
(H_0)$ \cite{PG}. One can then define the function ${\cal
Y}(H_0):=Y[H_0,\pi(H_0)]$. In these circumstances, it suffices that
the system satisfies, e.g., one of the following generic sets of
hypotheses to prove that the physical length uncertainty is strictly
positive.

i) We first assume that the function ${\cal Y}(H_0)$ is strictly
monotonic, namely $d{\cal Y}/dH_0\neq 0$, so that it provides a
one-to-one map. Then, via the spectral theorem, the eigenstates of
the operators ${\cal Y}$ and $H_0$ coincide, and the demand $\Delta
Y\!\!=\!\Delta {\cal Y}\!=0$ implies that $\Delta H_0\!=\!0$. The
fourth Heisenberg relation leads to $\Delta T \rightarrow \infty$.
Let us then prove that the third term in Eq. (\ref{unc}) does not
vanish when $\Delta H_0$ tends to zero. Expanding ${\cal Y}$ around
the mean value of $H_0$ \cite{nota}, we find
\begin{equation}
(\Delta {\cal Y})^2=\langle \widehat{\cal Y}^{2}-\langle
\widehat{\cal Y} \rangle^{2}\rangle \approx \left(
\left.\frac{d{\cal Y}}{dH_0}\right|_{\langle\widehat{H}_0\rangle}
\Delta H_0\right)^2,\end{equation} \begin{equation}\label{limit}
\lim_{\Delta H_0\rightarrow 0}\!2\Delta T\Delta {\cal Y}\!\geq \!
\lim_{\Delta H_0\rightarrow 0} \frac{\Delta{\cal Y}}{\Delta H_0}\!
=\! \left| \left.\frac{d{\cal Y}}{dH_0}\right|_{\langle
\widehat{H}_0\rangle}\right|\!\neq 0.
\end{equation}
We hence conclude that the physical length uncertainty cannot vanish
in this case.

ii) We suppose instead that ${\cal Y}(H_0)$ is positive and, for
large pseudo energies, grows at least like $H_0$ multiplied by a
constant. We analyze first the case in which ${\cal Y}$ is strictly
positive. Since $\langle \widehat{\,Y}\rangle=\langle \widehat{\cal
Y}\rangle$ is then different from zero, the vanishing of the second
term in Eq. (\ref{unc}) requires $\Delta T=0$. So, the fourth
Heisenberg relation implies that $\Delta H_0 \rightarrow \infty$.
Let us consider again the third term in Eq. (\ref{unc}). Our
condition on the behavior of ${\cal Y}$ for large $H_0$ can be
rephrased by saying that $\lim_{H_0\rightarrow \infty} ({\cal
Y}/H_0)>r$ for a certain number $r>0$. As a consequence, one can see
that $\lim_{\Delta H_0\rightarrow \infty} (\Delta{\cal Y}/\Delta
H_0)>r$. Therefore, the product $\Delta T\Delta Y=\Delta T \Delta
{\cal Y}$ cannot vanish when $\Delta H_0$ tends to infinity, and the
physical length uncertainty is strictly positive. On the other hand,
in the case that ${\cal Y}$ can also take the zero value, $\langle
\widehat{\,Y} \rangle=\langle \widehat{\cal Y}\rangle$ may
occasionally vanish, but this may only happen if the quantum state
is in the kernel of the operator $\widehat{\cal Y}$. We then
introduce the additional assumption that this kernel is formed
exclusively by the eigenvectors corresponding to a unique eigenvalue
$\overline{H}_0$ of $\widehat{H}_0$, a result that holds when ${\cal
Y}(H_0)$ vanishes only at that value of the pseudo energy. If the
system approaches such an eigenvector, the uncertainty of $H_0$
tends to zero and $\Delta T \rightarrow \infty$. Assuming finally
that $\left.(d{\cal Y}/dH_0) \right|_{{\overline H}_0} \!\neq 0$,
one arrives at the same conclusion about the third term in Eq.
(\ref{unc}) that was obtained in inequality (\ref{limit})
\cite{nota}. Therefore, under this set of hypotheses, it is
impossible to achieve an infinite resolution in the physical length.

An important class of DSR theories in which the positivity of ${\cal
Y}(H_0)$ is satisfied when $s_T$ is represented by an explicitly
time-dependent operator is when the physical energy does not depend
on the pseudo momentum, i.e., when the function $g$ depends only on
$H_0$. In this case, \[M=\frac{\partial_\pi g}{J}=0,\quad
N=\frac{\partial_{H_0}g}{J}=\frac{1}{\partial_\pi f},\quad
Y=\frac{H_0^{\prime}}{\partial_{\pi} f}.\]  As a consequence, ${\cal
Y}(H_0)$ is non-zero, because both the map $U$ and $H_0(\pi)$ are
invertible by assumption (this guarantees that $\partial_{\pi}f\neq
0$ and $H_0^{\prime}\neq 0$). Since ${\cal Y}(H_0)$ has a definite
sign, and $\partial_\pi f\approx 1$ in the sector of small pseudo
energy-momentum, in the standard situation with a pseudo energy that
increases with $\pi$ in that sector we conclude that ${\cal Y}(H_0)$
is strictly positive \cite{nota2}.

In conclusion, a non-vanishing uncertainty generically affects the
physical length in the perturbative quantization of the system. The
above discussion can also be applied to the study of the physical
time uncertainty considered in Ref. \cite{PG}. All the hypotheses
can be easily generalized to that case with the due substitution of
${\cal Y}$ by the function ${\cal V}$ defined in that reference.

\section{Physical Position Uncertainty: Non-Perturbative Case}
\setcounter{equation}{0}

We turn now to the analysis of the physical length uncertainty when
one adopts what we have called a non-perturbative quantization,
i.e., when the quantum evolution is described in terms of the
physical time.

In principle, one can always construct a non-per\-turbative quantum
theory (in the sense indicated above) starting with the perturbative
one, which has been assumed to exist. Employing the spectral
decomposition of the pseudo momentum $\pi$ and recalling that
$H_0=H_0(\pi)$, one can define the physical Hamiltonian
$H=g(H_0,\pi)$ as an operator. The parameter of the evolution
generated by this Hamiltonian can be identified with the physical
time $t$. By contrast, the background time gets now promoted to an
operator. This fact changes the expression of the observable
$\widehat{L}_{\eta}$ when regarded as an explicitly time dependent
operator. From Eqs. (\ref{X}) and (\ref{x}), one gets
\begin{eqnarray}
\label{Xt}\widehat{L}_{\eta}^{[2]}&=&\widehat{M}^{[2]}(H_0,\pi)\,t+
\widehat{R}_{t,\eta}^{[2]},\\ \nonumber\\
\widehat{R}_{t,\eta}^{[2]}&:=&\frac{1}{2}\left(
\widehat{N}^{[2]}(H_0,\pi)\widehat{S}_t+
\widehat{S}_t\widehat{N}^{[2]}(H_0,\pi)\right)
\nonumber\\
&+&\frac{1}{2}\left(
\widehat{O}_{\eta}^{[2]}(H_0,\pi)\widehat{s}_0+\widehat{s}_0
\widehat{O}_{\eta}^{[2]}(H_0,\pi)\right), \label{Ct}\end{eqnarray}
where \begin{eqnarray} \label{At} M^{[2]}&:=& \frac{\partial_{\pi}
g}{\partial_{\pi} f},\quad
N^{[2]}:=\frac{1}{\partial_{\pi} f}\\
O^{[2]}_{\eta}&:=&\eta \frac{\partial_{H_0}g-J} {\pi\,J}
-\frac{\partial_{\pi}g\,\partial_{H_0}f}{\pi\,J\,\partial_{\pi}f }.
\end{eqnarray}

The analysis is parallel to that followed in Sec. III and Sec. IV,
with the caveat that $s_t:=\pi_jq^j$ [and therefore
$S_t:=(s_t-s_0)/\pi$] must now be considered a variable that evolves
in the physical time $t$, rather than in the background time. In
particular, by extracting explicitly all the time dependence of
$s_t$ when defining its operator counterpart, one arrives at
\begin{equation}
\label{Xt2}\widehat{L}_{\eta}^{[2]}\!=\!\widehat{\,Y}^{[2]}(H_0,\pi)\,t+
\widehat{Z}_{\eta}^{[2]}(H_0,\pi,s_0),\end{equation} with
\begin{eqnarray}
\widehat{\,Y}^{[2]}(H_0,\pi)\!&=&\! \left(\widehat{H_0^{\prime}}
\widehat{\partial_{H_0} g}+\widehat{\partial_{\pi} g}\right)\!
\widehat{N}^{[2]}(H_0,\pi)\nonumber\\
&+&\widehat{M}^{[2]}(H_0,\pi),\label{Vt}\\
\label{Wt}\hspace*{-.5cm}\widehat{Z}_{\eta}^{[2]}(H_0,\pi,s_0)\!&=&
\!\frac{
\widehat{O}_{\eta}^{[2]}(H_0,\pi)\widehat{s}_0+\widehat{s}_0
\widehat{O}_{\eta}^{[2]}(H_0,\pi)}{2}.
\end{eqnarray}
Here, the observable $\widehat{s}_0$ represents the value of $s_t$
at the initial physical time, which is a constant of motion.

In order to calculate the physical length uncertainty, one has to
average now over the time parameter $t$, instead of averaging over
$T$, as we did in Eq. (\ref{uncert}). This leads to
\begin{eqnarray}
\left(\Delta
L_{\eta}^{[2]}\right)^2&=&\left[\Delta\left(Y^{[2]}\,\overline{t}+
Z_{\eta}^{[2]}\right)\right]^2 +\left(\langle
\widehat{\,Y}^{[2]}\rangle \Delta
t\right)^2\nonumber\\&+&\left(\Delta t\Delta
Y^{[2]}\right)^2,\label{unc2}
\end{eqnarray}
where $\overline{t}$ and $\Delta t$ are the mean value and the
uncertainty of the distribution deduced for the parameter $t$ by
analyzing the evolution of the probability densities of observables
in our quantum state. Obviously, the time uncertainty satisfies the
fourth Heisenberg relation $\Delta t\Delta H\geq 1/2$.

Notice that the physical length uncertainty is again given by the
sum of three positive terms. The analysis of the previous section
can be easily extended to the case considered here. From the
behavior of $\Delta L_{\eta}^{[2]}$ at large times we conclude that
$\Delta Y^{[2]}$ must vanish. Moreover, taking into account the
assumption that the function $H_0(\pi)$ be invertible, remembering
that $H=g(H_0,\pi)$, and using the implicit function theorem, it is
possible to define $Y^{[2]}$ as a function of only $H$ -that we
denote ${\cal Y}^{[2]}(H)$- provided that
$H_0^{\prime}\,\partial_{H_0} g+\partial_{\pi} g\neq 0$. One can
then introduce the same two sets of hypotheses that were discussed
in Sec. IV, but with the role of ${\cal Y}(H_0)$ played by ${\cal
Y}^{[2]}(H)$. In this way one concludes that, under quite generic
assumptions, an infinite resolution cannot be reached for the
physical length in a non-perturbative quantization of the system
constructed from the perturbative quantum theory.

Finally, we want to comment on the possibility that the system might
admit a different non-perturbative quantization (with evolution
still generated by the physical Hamiltonian) in which the
canonically conjugate physical variables $(X,p)$ were promoted to
explicitly time-independent operators and such that the quantum
spectrum of the physical momentum $p$ were contained in its
corresponding classical domain. This is non-trivial in general, and
the viability of such a quantization cannot be taken for granted
starting from the only assumption of the existence of a perturbative
quantum description with the properties that we have discussed. From
Eq. (\ref{projection}), we see that a situation in which this
possibility is realized is when the physical energy does not depend
on the pseudo momentum, $\partial_{\pi}g=0$. In this case (which
includes the example of the Einstein-Rosen waves), the physical
position $X$ is independent of the background time. It may then be
promoted to an operator that does not display any explicit time
dependence, in terms of those for $\pi^i$ and for the background
coordinates $q^i$, the latter evolving only implicitly in the time
parameter. Strictly speaking, nonetheless, the discussion presented
in the paragraphs above cannot be applied in these circumstances
because, with such an operator representation, $Y^{[2]}(H_0,\pi)$
must be identified with $M^{[2]}(H_0,\pi)$, the latter being
identically zero when so is $\partial_{\pi}g$ [see Eqs. (\ref{Xt})
and (\ref{At})]. This vanishing invalidates the sets of hypotheses
under which our study was carried out.

When a non-perturbative quantization with those cha\-racteristics
exists, the Heisenberg uncertainty principle implies that $\Delta X
\Delta p \geq 1/2$. As a consequence, the resolution in the physical
position is limited if and only if the physical momentum is bounded
from above. This happens in DSR1 and DSR2 theories, but not in DSR3.
The same phenomenon occurs with the physical length if it is
determined by the difference of two uncorrelated position
observables. In conclusion, we see that the emergence of a minimum
uncertainty in the physical length is unavoidable non-perturbatively
as well as perturbatively, except perhaps for DSR3 theories that
admit a non-perturbative quantization in which $X$ can be
represented as an explicitly time-independent observable.

\section{First Order Corrections in the Perturbative Case}
\setcounter{equation}{0}

In this section we will study the physical length uncertainty that
arises in the perturbative quantization when the operator
$\widehat{L}_{\eta}$ is approximated up to first order corrections
in the energy. To obtain this approximation, we expand the functions
$f$ and $g$ (which we suppose smooth) in the variables $H_0$ and
$\pi$ around their minimum values. Motivated by the case of free
particles in special relativity, we assume that the minimum
magnitude of the pseudo momentum is zero, whereas the minimum of the
pseudo energy $\mu$ will be just non-negative \cite{PG}. We then
denote ${\cal H}_0:=H_0-\mu$ and keep only up to quadratic terms in
${\cal H}_0$ and $\pi$ in the expansions of the two functions; this
truncation will suffice for our purposes. In addition, we suppose
that $\mu$ is small compared with the invariant energy/momentum
scale of the DSR theory, so that the leading terms in the region of
expansion are $f(H_0,\pi)\approx\pi$ and $g(H_0,\pi)\approx H_0$
(because the map $U$ determined by $f$ and $g$ must approach the
identity in the low energy-momentum sector).

From Eq. (\ref{A}), one then gets
\begin{eqnarray}\label{ABFO}
M(H_0,\pi)&\approx&\left.(\partial_{H_0}\partial_{\pi}g)\right|_0{\cal
H}_0+\left.(\partial^{2}_{\pi}g)\right|_0\pi ,\nonumber\\
N(H_0,\pi)&\approx&
1-\left.(\partial_{H_0}\partial_{\pi}f)\right|_0{\cal
H}_0-\left.(\partial^{2}_{\pi}f)\right|_0\pi,
\end{eqnarray}
where the symbol $\,\!|_0$ represents evaluation at ${\cal
H}_0=\pi=0$. Substituting these results and the expression
$H_0(\pi)$ of the background Hamiltonian in Eqs. (\ref{V}) and
(\ref{W}) [and recalling definitions (\ref{A})], we deduce the first
order approximation for the operators $\widehat{Y}$ and
$\widehat{Z}_{\eta}$. An extrapolation of the situation found in
special relativity \cite{PG} leads us to consider the following
cases.

1) \emph{Massive} case: $\mu\neq0$, with
$H_0^{\prime}|_{\pi=0}=0$.\\
We obtain $H_0(\pi)\approx\mu+b\pi^2$, where
$2b:=H_0^{\prime\prime}|_{\pi=0}$. Assuming that $b>0$, we have that
$\pi\approx \sqrt{{\cal H}_0/b}$. Thus, we can neglect terms
proportional to ${\cal H}_0$ with respect to those linear in $\pi$.
In this way, one finds
\begin{eqnarray}
\label{massV}\widehat{Y}&\approx&\left[2b+\left.
(\partial^{2}_{\pi}g)\right|_0\,
\right]\widehat{\pi} ,\\
\label{massW}\widehat{Z}_{\eta}&\approx&-\eta\,
\left.(\partial^{2}_{\pi}f)\right|_0\widehat{s}_0,
\end{eqnarray}
where we have employed that $s_0=\left.\pi_j q^j\right|_{T=0}$ is of
the same order as $\pi$.

The function ${\cal Y}$, defined in Sec. IV, is given in this
approximation by the classical analog of Eq. (\ref{massV}) with
$\pi=\sqrt{{\cal H}_0/b}$. The resulting function is strictly
monotonic in $H_0$ if the constant coefficient $2b+\left.
(\partial^{2}_{\pi}g)\right|_0$ does not vanish, as it must happen
if our truncation provides indeed the first order approximation.
Therefore, the first set of hypotheses considered in Sec. IV is
applicable in this case, leading us to the conclusion that it is
impossible to achieve an infinite resolution in the physical length.

2) \emph{Massless} case: $\mu\!=\!0$, with
$H_0^{\prime}|_{\pi=0}\!=k\neq 0$.\\
Now ${\cal H}_0=H_0\approx k\pi$, so that corrections proportional
to either $H_0$ or $\pi$ are of the same order. We then arrive at
\begin{eqnarray}
\label{mlV}\widehat{Y}&\approx & k+
\left[\frac{2b}{k}-\left.(\partial^{2}_{\pi}f)\right|_0 -k\left.
(\partial_{H_0}\partial_{\pi}f)\right|_0 \right.\nonumber\\
&&+\left.\frac{\left.(\partial^{2}_{\pi}g)\right|_0}{k}
+\left.(\partial_{H_0}\partial_{\pi}g)\right|_0
\right]\widehat{H}_0 ,\\
\label{mlW}\widehat{Z}_{\eta}&\approx &-\eta \left[
k\left.(\partial_{H_0}\partial_{\pi}f)\right|_0
+\left.(\partial^{2}_{\pi}f)\right|_0\right] \widehat{s}_0.
\nonumber
\end{eqnarray}
The constant $b$ is defined as in the \emph{massive} case. The
next-to-leading order approximation to the function ${\cal Y}$ is
thus given by the classical counterpart of Eq. (\ref{mlV}). Again,
provided that the constant coefficient of the first order correction
in $H_0$ differs from zero, the function ${\cal Y}$ is strictly
monotonic. The physical length uncertainty is hence greater than
zero in this approximation.

\section{First Order Corrections: Behavior at Large Times}
\setcounter{equation}{0}

In this section, we will analyze in more detail the physical length
uncertainty in the perturbative quantization for the \emph{massless}
case adopting the next-to-leading order approximation for low
energies. We will pay a special attention to the behavior displayed
at large values of the background time. We will show that this
behavior is of the kind that was first discussed by Salecker and
Wigner \cite{SalWig}. Since a similar study was not considered in
Ref. \cite{PG} for the physical time uncertainty, we will carry out
our analysis in a way that is also valid for it.

From the results of Ref. \cite{PG} and our comments above, the
physical time lapse $t_{\eta}$ is affected in the perturbative
quantization by the uncertainty:
\begin{equation}\label{unct}
(\Delta t_{\eta})^2\!=\![\Delta(V\overline{T}+W_{\eta})]^2 +\langle
\widehat{\,V}\rangle^2(\Delta T)^2+(\Delta T\Delta V)^2,
\end{equation}
where the operators $\widehat{V}$ and $\widehat{W}_{\eta}$ have
these expressions in the considered approximation for the
\emph{massless} case:
\begin{eqnarray}
\label{mlVt}\widehat{V}&\approx & 1+
\bigg[k\left.(\partial^{2}_{H_0}f)\right|_0 +\left.
(\partial_{H_0}\partial_{\pi}f)\right|_0 \nonumber\\
&&-\left.(\partial^{2}_{H_0}g)\right|_0
-\frac{\left.(\partial_{H_0}\partial_{\pi}g)\right|_0}{k}\,
\bigg]\,\widehat{H}_0 ,\\
\label{mlWt}\widehat{W}_{\eta}&\approx &\eta \left[
\left.(\partial_{H_0}\partial_{\pi}f)\right|_0
+k\left.(\partial^{2}_{H_0}f)\right|_0\right] \widehat{s}_0.
\nonumber
\end{eqnarray}
We then introduce the notation
$\{L_{\alpha,\eta}\}:=\{t_{\eta},L_{\eta}\}$,
$\{Y_{\alpha}\}:=\{V,Y\}$, and
$\{Z_{\alpha,\eta}\}:=\{W_{\eta},Z_{\eta}\}$ to describe
simultaneously the formulas for the physical time and length
uncertainties. Let us emphasize that $\alpha=0,1$ is just an
abstract subscript notation.

After a trivial elaboration, we can rewrite Eqs. (\ref{unc}) and
(\ref{unct}) as
\begin{eqnarray}\label{unca}
(\Delta L_{\alpha,\eta})^2&=&\overline{T}^2(\Delta
Y_{\alpha})^2+(\Delta Z_{\alpha,\eta})^2 +\overline{T}\,{\rm
cov}(\widehat{Y}_{\alpha},\widehat{Z}_{\alpha,\eta})
\nonumber\\
&&+\langle\widehat{Y}_{\alpha}\rangle^2(\Delta T)^2+(\Delta T\Delta
Y_{\alpha})^2.
\end{eqnarray}
No sum over ${\alpha}$ is implied and
\begin{equation}{\rm cov}(\widehat{Y}_{\alpha},\widehat{Z}_{\alpha,\eta})
:=\langle\widehat{Y}_{\alpha}\widehat{Z}_{\alpha,\eta}+
\widehat{Z}_{\alpha,\eta}\widehat{Y}_{\alpha}\rangle-2
\langle\widehat{Y}_{\alpha}\rangle\langle\widehat{Z}_{\alpha,\eta}\rangle.
\end{equation}
In addition, in the studied approximation for the \emph{massless}
case, we can write the operators $\widehat{Y}_{\alpha}$ and
$\widehat{Z}_{\alpha,\eta}$ in the form
$\widehat{Y}_{\alpha}=\kappa_{\alpha}+\lambda_{\alpha}\widehat{H}_0/
E_P$ and
$\widehat{Z}_{\alpha,\eta}=\eta\delta_{\alpha}\widehat{s}_0/E_P$
[see Eqs. (\ref{mlV}) and (\ref{mlVt})], where $E_P$ is the Planck
energy ($E_P=1/\sqrt{G}$ in our units, $G$ being Newton constant),
$\lambda_{\alpha}$ and $\delta_{\alpha}$ are appropriate constant
coefficients that differ from zero, $\kappa_0:=1$, and
$\kappa_1:=k=\left.H_0^{\prime}\right|_{\pi=0}$.

The last term in Eq. (\ref{unca}) is then
\begin{equation}(\Delta T\Delta Y_{\alpha})^2= \frac{\lambda_{\alpha}^2
(\Delta T\Delta H_0)^2}{E_P^{2}}\geq\frac{\lambda_{\alpha}^2
l_P^2}{4}.\end{equation} In the last step, we have used the fourth
Heisenberg relation for the background time and energy, and
introduced the Planck length $l_P=1/E_P$ (in our units). Recalling
that the other contributions to the physical uncertainty are
positive, we conclude that $\Delta L_{\alpha,\eta}\geq
|\lambda_{\alpha}|l_P/2$. Therefore, we see that the uncertainty in
both the physical time lapse and the physical length is bounded from
below by a contribution of quantum gravitational origin that is of
the order of the Planck length \cite{Luis,Pad,unc}.

From the rest of contributions to the physical uncertainty
(\ref{unca}), one gets in a similar way the bound
\begin{eqnarray}
\label{F}(\Delta
L_{\alpha,\eta})^2&>&\lambda_{\alpha}^2\overline{T}^2\frac{(\Delta
H_0)^2}{E_P^2}+\frac{\langle\widehat{Y}_{\alpha}\rangle^2}
{4(\Delta H_0)^{2}}+(\Delta Z_{\alpha,\eta})^2\nonumber\\
&&+\;\overline{T}{\rm cov}
(\widehat{Y}_{\alpha},\widehat{Z}_{\alpha,\eta}).
\end{eqnarray}
The r.h.s. of this inequality can be regarded as a function of the
uncertainty in the background energy $\Delta H_0$, once the
next-to-leading order expressions for the operators
$\widehat{Y}_{\alpha}$ and $\widehat{Z}_{\alpha,\eta}$ have been
substituted. Hence, for uncertainties $\Delta H_0$ in a certain
interval, one can deduce a more general bound for $\Delta
L_{\alpha,\eta}$ by minimizing that function. The extrema can be
deduced by imposing the vanishing of the first derivative with
respect to $\Delta H_0$:
\begin{eqnarray}
\label{dF}0&=&2\lambda_{\alpha}^2\overline{T}^2\frac{(\Delta
H_0)^4}{E_P^2}-\frac{\langle\widehat{Y}_{\alpha}\rangle^2}{2}
+(\Delta H_0)^3\partial_{\Delta}(\Delta Z_{\alpha,\eta})^2 \nonumber\\
&+&\!\frac{\Delta H_0\partial_{\Delta}(\langle\widehat{Y}_{\alpha}
\rangle^2)}{4} +\overline{T}(\Delta H_0)^3\partial_{\Delta}{\rm
cov}(\widehat{Y}_{\alpha}, \widehat{Z}_{\alpha,\eta}).
\end{eqnarray}
Here, we have introduced the notation $\partial_\Delta$ to denote
the derivative with respect to $\Delta H_0$.

Provided that $\langle\widehat{Y}_{\alpha}\rangle$ can be considered
independent of both $\Delta H_0$ and the (mean value of the)
background time $\overline{T}$, the first two terms in the r.h.s. of
Eq. (\ref{F}) are in fact the kind of contributions that lead to the
emergence of a minimum uncertainty of the Salecker and Wigner type
(see Appendix A for details) \cite{SalWig,BA}. Namely, we get a
contribution that is linear in $(\Delta H_0)^2$ and another one that
is proportional to its inverse. If these two terms were the only
ones that appeared in our equations, an analysis similar to the
standard one for Salecker-Wigner devices would prove that the bound
for $\Delta L_{\alpha,\eta}$ reaches its minimum at a value of
$\Delta H_0$ that scales with the background time like $\Delta
H_0^{min} \propto1/\sqrt{\overline{T}}$, whereas the lower bound
obtained for the physical uncertainty at $\Delta H_0^{min}$
increases in time like $\sqrt{\overline{T}}$.

Motivated by these remarks, we will now show that, at least in the
region of small $\Delta H_0$ and for large values of the background
time $\overline{T}$, the terms in Eqs. (\ref{F}) and (\ref{dF})
other than the first two ones do not invalidate the above
conclusions about the existence of a (local) minimum and its
associated bound. The restriction to small values of $\Delta H_0$ is
natural in the context of the low-energy approximation that we are
discussing. Moreover, for unboundedly large times $\overline{T}$,
the sector of vanishingly small values of $\Delta H_0$ contains the
relevant region for the analysis of the Salecker-Wigner bound on the
uncertainty, i.e. the region around the minimum $\Delta
H_0^{min}\propto1/\sqrt{\overline{T}}$.

In this sector of background energy uncertainties and time, one can
demonstrate that a set of sufficient conditions to deduce a
Salecker-Wigner behavior are:
\begin{eqnarray}\label{limits}
\mbox{a) }&\lim_{\Delta
H_0\rightarrow0}&\langle\widehat{Y}_{\alpha}\rangle^2=c^{(1)}_{\alpha},
\nonumber\\
\mbox{b) }&\lim_{\Delta H_0\rightarrow0}&(\Delta H_0)^2(\Delta
Z_{\alpha,\eta})^2=c^{(2)}_{\alpha},\nonumber\\
\mbox{c) }&\lim_{\Delta H_0\rightarrow0}&(\Delta
H_0)^3\partial_\Delta
(\Delta Z_{\alpha,\eta})^2=c^{(3)}_{\alpha},\nonumber\\
\mbox{d) }&\lim_{\Delta H_0\rightarrow0}&\Delta
H_0\partial_\Delta\langle
\widehat{Y}_{\alpha}\rangle^2=0,\nonumber\\
\mbox{e) }&\lim_{\Delta H_0\rightarrow0}&{\rm cov}
(\widehat Y_{\alpha},\widehat Z_{\alpha,\eta})=0,\nonumber\\
\mbox{f) }&\lim_{\Delta H_0\rightarrow0}&\Delta H_0\partial_\Delta
{\rm cov} (\widehat Y_{\alpha},\widehat Z_{\alpha,\eta})=0,
\end{eqnarray}
where $c^{(n)}_{\alpha},\mbox{ n=1,2,3}$, are constants (with
$c^{(1)}_{\alpha}-2c^{(3)}_{\alpha}\neq 0$ and
$c^{(1)}_{\alpha}+2c^{(2)}_{\alpha}-c^{(3)}_{\alpha}\neq 0$).
Conditions a), b), and c) allow one to absorb the third term in the
r.h.s. of Eqs. (\ref{F}) and (\ref{dF}) just as a modification to
$\langle\widehat{Y}_{\alpha}\rangle^2$ and treat this (square)
expectation value as a constant when calculating the value of our
function around its extrema in the region $\Delta H_0\ll 1$. In such
a calculation and for sufficiently large background times,
conditions d), e), and f) guarantee that all but the first three
terms in Eqs. (\ref{F}) and (\ref{dF}) can be neglected.

Taking into account that $\widehat{Z}_{\alpha,\eta}$ vanishes when
$\eta=0$, the only non-trivial requirements in that case are
conditions a) and d). Regardless of the value of $\eta$, we prove in
Appendix B that all the above conditions are satisfied at least for
quantum states that are described by gaussian wave packets
\cite{nota3}. Since we are assuming the feasibility of a
(perturbative) quantization with canonical variables given by the
background flat spatial coordinates and the pseudo momentum, and in
addition we have focused our discussion on free systems, it seems
reasonable to suppose that such states exist and provide the analog
of classical particles in our quantum theory. Besides, the
limitation to wave packets is already present in the deduction of
the Salecker-Wigner bound for the spacetime uncertainty (in order to
justify the assumption that the position and momentum operators have
vanishing covariance) \cite{BA}. So, it is natural to incorporate
the same restriction to our analysis.

Substituting the values of the constants $c_n$ computed in Appendix
B (under the simplifying assumption of only one spatial dimension),
one obtains the following bounds for large background times from the
corresponding minima in the region $\Delta H_0\ll 1$:
\begin{equation}\label{root} (\Delta
L_{\alpha,\eta})^2>d_{\alpha,\eta}\;l_p\overline{T},\end{equation}
where
\begin{equation}
d_{\alpha,\eta}=\lambda_{\alpha}\!\left[\eta k^2
\frac{\delta_{\alpha}^2}{E_P^{2}}\nu^2+
\left(\kappa_{\alpha}+k\frac{\lambda_{\alpha}}{E_P}|\nu|\right)^2
\right]^{\frac{1}{2}}.
\end{equation}
Here, $\nu$ denotes the expectation value of the pseudo momentum.

In conclusion, in the perturbative quantization of free
\emph{massless} systems in DSR theories and within the low-energy
approximation, we have seen that the physical time and length
uncertainties are always bounded from below by a quantum
gravitational contribution of the order of the Planck length, while
for large values of the background time the uncertainties increase
like $\sqrt{l_P\overline{T}}$ (at least for wave packets), just like
in Salecker-Wigner devices.

\section{Conclusion}

In this work, we have analyzed the emergence of a minimum
non-vanishing length uncertainty in the framework of a gravity's
rainbow formalism, derived from a dual realization of DSR theories
in spacetime. This realization leads to a set of spacetime
coordinates that are canonically conjugate to the physical energy
and momentum. Therefore, the transformation from the background
energy-momentum and spacetime coordinates (also called pseudo
variables) to those that we consider as physical is provided by a
canonical transformation. In particular, the physical spacetime
variables are linear in the background ones, but in general depend
nonlinearly on the pseudo energy and momentum of the particle.

We have specialized our analysis to systems that admit a Hamiltonian
formulation, with the energy determined by the value of the
Hamiltonian, and concentrated our attention on the case of a free
dynamics, motivated by the consideration of DSR theories as
(effective) descriptions of free particles in special relativity
modified by gravity. In these free systems, the background
Hamiltonian is a function of only the (magnitude of the) pseudo
momentum. We have studied the behavior of the physical position,
understanding as such the scalar obtained by projecting the physical
position vector in the momentum direction. More specifically, we
have investigated the quantum uncertainty that affects the physical
length, defined by the difference between this physical position and
the initial value of the position, either in the background or in
the physical variables of the system. This study has been carried
out in two possible quantization schemes, referred as perturbative
and non-perturbative quantizations.

The perturbative approach corresponds to a quantization in which the
evolution is generated by the background Hamiltonian, so that the
background time $T$ plays the role of evolution parameter. We have
assumed that a quantum theory of this kind is feasible. In this
quantization, the physical time and length are represented by
genuine operators that depend explicitly on the time parameter. We
have been able to generalize the analysis of Ref. \cite{PG} for the
physical time uncertainty, and prove that the uncertainty in the
physical length is also strictly positive in this approach.

Rigorously speaking, we have demonstrated this positivity under two
different sets of generic assumptions. Both sets contain the more
than reasonable hypothesis that the considered quantum state has a
finite expectation value of the background energy,
$\langle\widehat{H}_0\rangle<\infty$. Besides, the two sets include
an assumption about the functional dependence of the background
energy on the pseudo momentum, namely, that the function
$H_0=H_0(\pi)$ be invertible. The rest of hypotheses concern the
detailed form of the DSR theory, and more concretely the properties
of the function ${\cal Y}(H_0):=Y[H_0,\pi(H_0)]$ introduced in Sec.
IV.

One set of assumptions requires this function to be strictly
monotonic, i.e. ${\cal Y}^{\prime}(H_0)\neq0$ for all values of
$H_0$.

The other set involves several requirements. The most important ones
are: i) the positivity of ${\cal Y}$, ${\cal Y}\geq0$; and ii) a
linear or faster increase of ${\cal Y}$ with $H_0$ at infinity,
$\lim_{H_0\rightarrow\infty}({\cal Y}/H_0)>r$ for a certain constant
$r>0$. In addition, it is demanded that: iiia) the kernel of ${\cal
Y}$ be empty, or either iiib) this kernel consist of a single point
$\overline{H}_0$ where the derivative of ${\cal Y}$ does not vanish,
$\left.(d{\cal Y}/dH_0)\right|_{\overline{H}_0}\neq0$.

In the non-perturbative approach, on the other hand, the evolution
is generated by the physical Hamiltonian, and the physical time $t$
is identified with the evolution parameter. Starting with the
perturbative quantization that we have assumed to exist, it is in
general possible to construct a non-perturbative quantum theory of
this kind, in which the physical length is represented by an
operator that depends explicitly on the time parameter $t$. We have
proved that the quantum uncertainty in this operator is strictly
positive under similar sets of assumptions to those discussed for
the case of the perturbative quantization. Therefore, it is again
impossible to reach an infinite resolution in the physical length.

It might also happen that the system admits a different
non-perturbative quantization in which the evolution is indeed
generated by the physical Hamiltonian, but the physical position
variable gets promoted to an operator that is explicitly independent
of time and canonically conjugate to the operator which represents
the magnitude of the physical momentum. In general, the existence of
such a quantum theory is not granted from the sole assumption of the
viability of the perturbative quantization. Supposing besides that
the quantum spectrum of the physical momentum is contained in its
classical domain, Heisenberg principle implies that the uncertainty
in the physical position can be made to vanish only if the physical
momentum is not bounded from above. The same result holds for the
physical length if it is determined by the difference of two
uncorrelated physical positions.

The existence of an upper bound for the physical momentum, with the
consequent limit in the spatial resolution, occurs only in the DSR1
and DSR2 families, but not in DSR3 theories. Remarkably, for such
theories the physical time uncertainty is always bounded away from
zero in the non-perturbative quantum theory \cite{PG}. As a result,
it is never possible to reach an infinite resolution, both in the
physical time and position, in the non-perturbative quantization of
Hamiltonian free systems within the context of DSR theories.

Finally, we have also analyzed the uncertainty in the perturbative
quantization when the operator corresponding to the physical length
is approximated up to first order corrections in the energy. The
study has lend support to the conclusion that this uncertainty is
generically greater than zero. Special attention has been paid to
the \emph{massless} case, in which the background energy is
proportional to the magnitude of the pseudo momentum in the
considered approximation. We have proved that, in that case, the
uncertainty is always bounded by a quantity of the order of the
Planck length. This bound can be interpreted as a contribution of
quantum gravitational origin. In addition we have proved that, in
the low-energy regime and for large values of the background time,
the uncertainties in the physical time and length admit lower bounds
that increase with the square root of time. This is precisely the
kind of behavior that was suggested by Salecker and Wigner for
spacetime measurements made with quantum devices.

\acknowledgments

The authors want to thank B. Barcel\'o and J. Cort\'es for helpful
conversations. P.G. acknowledges CSIC and the European Social Fund
for the financial support provided by an I3P grant. This work was
supported by funds provided by the Spanish MEC-MCYT projects
BFM2002-04031-C02-02 and FIS2004-01912.

\section*{Appendix A: Salecker-Wigner Devices}
\setcounter{equation}{0}
\renewcommand{\theequation}{A.\arabic{equation}}

In this appendix we will briefly summarize the rationale of Salecker
and Wigner about the quantum uncertainty that is inherent to the
measurement of spacetime distances \cite{SalWig,BA}. The analysis
starts with the consideration of a measurement device, regarded as a
free system with mass $m$ and uncertainties in its initial position
and momentum $\Delta q$ and $\Delta \pi$. The (square) uncertainty
in its position at a later instant of time $t$ is
\begin{eqnarray}
[\Delta
q(t)]^2&=&\left[\Delta\!\left(q+\frac{t}{m}\pi\right)\!\right]^2
\nonumber\\
&=&(\Delta q)^2+\frac{t^2}{m^2}(\Delta \pi)^2+\frac{t}{m}{\rm
cov}(\widehat q,\widehat \pi),\nonumber
\end{eqnarray}
where ${\rm cov}(\widehat q,\widehat \pi\,)
:=\langle\widehat{q}\,\widehat{\pi}+\widehat{\pi}\,\widehat{q}\,
\rangle-2 \langle\widehat{q}\,\rangle\langle\widehat{\pi}\,\rangle.$
This expression gets simplified when the (initial) position and
momentum observables are not correlated. This occurs, for instance,
if the states of the system are plane waves modulated by a Gaussian.
In that case ${\rm cov}(\widehat{q},\widehat{\pi})=0$. Making use of
the fourth Heisenberg relation, one then obtains the inequality
\begin{equation}\label{SW}
[\Delta q(t)]^2\geq\frac{t^2}{m^2}(\Delta \pi)^2+\frac{1}{4(\Delta
\pi)^{2}}.
\end{equation}
The r.h.s. of this equation can be viewed as a function of $\Delta
\pi$. Its extrema can be determined by imposing the vanishing of the
first derivative: \[0=\frac{4t^2}{m^2}(\Delta \pi)^4- 1.\] The
minimum value of the uncertainty is hence reached at $\Delta
\pi^{min}=\sqrt{m/(2t)}$. Substituting this value in (\ref{SW}) one
gets a lower bound for the position uncertainty at the instant $t$:
\[\Delta q(t)\geq\sqrt{\frac{t}{m}}\,.\]

Therefore, the arguments of Salecker and Wigner imply that the
uncertainty increases with the square root of time.

\section*{Appendix B: Calculations for Wave Packets}
\setcounter{equation}{0}
\renewcommand{\theequation}{B.\arabic{equation}}

This appendix contains the calculation of the mean values,
uncertainties and covariance of the operators $\widehat{Y}_{\alpha}$
and $\widehat{Z}_{\alpha,\eta}$ introduced in Sec. VII, adopting the
next-to-leading order approximation for low energies and restricting
the quantum states to be gaussian wave packets (in the free quantum
theory with elementary variables given by the background spatial
coordinates and momenta). Moreover, in order to simplify our
calculations, we will carry out our analysis not in three, but just
in one spatial dimension. We do not expect this reduction to
qualitatively affect our results.

Explicitly, we will adopt a standard momentum representation in one
dimension, with wave packets given by the following wave functions
\cite{nota3}:
\[
\Psi(\pi_1)=\frac{1}{(2\Pi\sigma^2)^{1/4}}\,e^{-(\pi_1-\nu)^2/
(4\sigma^2)}e^{-i\mu \pi_1}.
\]
Here, $\nu:=\langle\widehat{\,\pi_1}\,\rangle$, $\sigma:=\Delta
\pi_1$, and $\mu:=\langle\widehat{ q^1}\rangle$, with $q^1$ being
the initial background position (we obviate its subscript $0$ to
simplify the notation). The number $\Pi$ is denoted in this appendix
with a capital Greek letter in order to distinguish it from the
magnitude of the pseudo momentum $\pi$. Besides, note that in one
dimension $\pi=|\pi_1|$.

From the functional form of the wave packets, it is clear that the
quantities that we want to compute will depend on the parameters
$\mu$, $\nu$, and $\sigma$. So, in order to calculate the limiting
values (\ref{limits}), we need to express the limit $\Delta
H_0\rightarrow 0$ in terms of those parameters. In the studied
approximation, $H_0=k \pi$ for the \emph{massless} case, and a
trivial calculation shows that the uncertainty $\Delta H_0$ for the
wave packets is given by
\begin{eqnarray}
(\Delta H_0)^2&=&k^2(\Delta\pi)^2=k^2\left(\sigma^2+
\nu^2-\langle\widehat{\pi}\rangle^2\right)\nonumber\\
&:=&G^2(\sigma,\nu),\label{Del}\\
\label{perf} \langle\widehat{\pi}\rangle&=&\!|\nu|\,{\rm
erf}\left(\frac{|\nu|}{\sqrt{2}\sigma}\right)\!
+\!\sqrt{\frac{2}{\Pi}}\,\sigma\,
e^{-\nu^2/(2\sigma^2)}\!.\end{eqnarray} It is worth emphasizing that
$\langle\widehat{\pi}\rangle$, the expectation value of the
magnitude of the pseudo momentum, differs in general from $\nu$. We
have introduced the error function
\[{\rm
erf}(x)=\frac{2}{\sqrt{\Pi}}\int_0^{x}\!dy\, e^{-y^2}, \mbox{ with}
\lim_{x\rightarrow\infty}{\rm erf}(x)=1.\]

From the above equations, we see that
$\langle\widehat{\pi}\rangle\approx |\nu|$ and $\Delta H_0\approx
k\sigma$ for small uncertainties $\Delta H_0$. Via the implicit
function theorem, we can then use the relation $\Delta
H_0=G(\sigma,\nu)$ ($G$ being the square root of $G^2$) to define
$\sigma$ as a function of $\Delta H_0$ in a neighborhood of the
origin of these quantities, provided that $\partial_\sigma G$ does
not vanish there. Actually, one has that $\lim_{\sigma\rightarrow 0}
\partial_\sigma G=k\neq0$. Therefore, one is allowed to replace the
limit $\Delta H_0\rightarrow 0$ with $\sigma\rightarrow 0$. In
addition, one can substitute the partial derivative with respect to
$\Delta H_0$ (i.e., $\partial_\Delta$) by
$\partial_{\Delta}\sigma\,\partial_ {\sigma}$, where
$\lim_{\sigma\rightarrow0}\partial_{\Delta}\sigma=1/k$. These
considerations lead to the results given in the rest of this
appendix, where we analyze simultaneously the cases of the physical
time and length uncertainties.

In the first order approximation for the {\it \emph{massless}} case,
the operators $\widehat{Y}_{\alpha}$ and $\widehat{Z}_{\alpha,\eta}$
adopt expressions of the form [see Eqs. (\ref{mlV}) and
(\ref{mlVt})]:
\begin{eqnarray}
\widehat{Y}_{\alpha}&=& \kappa_{\alpha}+k
\frac{\lambda_{\alpha}}{E_P}
\widehat{\pi},\nonumber\\
\widehat{Z}_{\alpha,\eta}&=& \eta\frac{\delta_{\alpha}}{E_P}
\widehat{s}_0=\eta\frac{\delta_{\alpha}}{2E_P}\left(\widehat{\,\pi_1}
\widehat{q^1}+\widehat{q^1}\widehat{\,\pi_1}\right),\nonumber
\end{eqnarray}
where $\lambda_{\alpha}$ and $\delta_{\alpha}$ are certain
non-vanishing constants, $\eta$ can take the values 0 or 1,
$\kappa_0=1$, and $\kappa_1=k$. We have employed that in this
approximation $\widehat{H}_0=k\widehat{\pi}$.

A straightforward calculation along the lines explained above shows
that for wave packets \[\lim_{\Delta
H_0\rightarrow0}\langle\widehat{Y}_{\alpha} \rangle^2
=\lim_{\sigma\rightarrow0}\langle\widehat{Y}_{\alpha}\rangle^2
=\left(\kappa_{\alpha}+k\frac{\lambda_{\alpha}}{E_P}
|\nu|\right)^2:=c_{\alpha}^{(1)}.\] In the same way, one finds
\begin{eqnarray}\Delta H_0\partial_\Delta\langle\widehat{Y}_{\alpha}
\rangle^2\!&=& \!2k\frac{\lambda_{\alpha}}{E_P}\!
\left(\kappa_{\alpha}+k
\frac{\lambda_{\alpha}}{E_P}\langle\widehat{\pi}\rangle\right)
\!\partial_{\sigma}\langle\widehat{\pi}\rangle\,\Delta
H_0\,\partial_{\Delta}\sigma,
\nonumber\\
\Delta H_0\,\partial_{\Delta}\sigma &=&\frac{\sigma^2+\nu^2-
\langle\widehat{\pi}\rangle^2}{\sigma-\langle\widehat{\pi}\rangle
\,\partial_{\sigma}\langle\widehat{\pi}\rangle}.\label{Depar}
\end{eqnarray}
From Eq. (\ref{perf}) one can check that
$\partial_{\sigma}\langle\widehat{\pi}\rangle$ tends fast enough to
zero when $\sigma\rightarrow 0$ ($\Delta H_0\rightarrow 0$) as to
guarantee that \[\lim_{\Delta H_0\rightarrow0}\Delta
H_0\partial_\Delta \langle\widehat{Y}_{\alpha}\rangle^2=0.\]

On the other hand, a similar computation leads to the following
uncertainty for the operator $\widehat{Z}_{\alpha,\eta}$
\begin{eqnarray}(\Delta Z_{\alpha,\eta})^2&=&
\eta\frac{\delta_{\alpha}^2}{E_P^{2}}
\left(\langle\widehat{s}_0^{\;2}\rangle-\langle
\widehat{s}_0\rangle^2\right)\nonumber\\
&=&\eta\frac{\delta_{\alpha}^2}{E_P^{2}}
\left(\frac{\nu^2}{4\sigma^2}+\mu^2
\sigma^2+\frac{1}{2}\right).\nonumber
\end{eqnarray}From this and Eqs. (\ref{Del}) and (\ref{Depar}),
it is not difficult to prove that
\begin{eqnarray}\lim_{\Delta H_0\rightarrow 0}(\Delta H_0)^2
(\Delta Z_{\alpha,\eta})^2&=&\eta\,k^2\frac{\delta_{\alpha}^2}
{4E_P^2}\,\nu^2:=c^{(2)}_{\alpha},
\nonumber\\
\lim_{\Delta H_0\rightarrow 0}(\Delta H_0)^3\partial_\Delta (\Delta
Z_{\alpha,\eta})^2&=&-\eta k^2\frac{\delta_{\alpha}^2}
{2E_P^2}\,\nu^2:=c^{(3)}_{\alpha}.\nonumber
\end{eqnarray}

Finally, the covariance of $\widehat{Y}_{\alpha}$ and
$\widehat{Z}_{\alpha,\eta}$ is given by \[{\rm
cov}(\widehat{Y}_{\alpha}, \widehat{Z}_{\alpha,\eta})= \eta
k\frac{\lambda_{\alpha}\delta_{\alpha}}{E_P^2} \left(\langle
\widehat{\pi}\widehat{s}_0+\widehat{s}_0\widehat{\pi}\rangle
-2\langle\widehat{\pi} \rangle\langle\widehat{s}_0\rangle\right) \]
which for wave packets gives \[ {\rm
cov}(\widehat{Y}_{\alpha},\widehat{Z}_{\alpha,\eta})= 2\eta
k\,\frac{\lambda_{\alpha}\delta_{\alpha}}{E_P^2}\,\mu \sigma^2\,
{\rm sign}(\nu)\,{\rm erf}
\left(\frac{|\nu|}{\sqrt{2}\sigma}\right).\] Therefore, one can
check that
\begin{eqnarray}&&\lim_{\Delta H_0\rightarrow 0}{\rm cov}
(\widehat{Y}_{\alpha},\widehat{Z}_{\alpha,\eta})=0,
\nonumber\\
&&\lim_{\Delta H_0\rightarrow 0}\Delta H_0
\partial_\Delta{\rm cov}(\widehat{Y}_{\alpha},
\widehat{Z}_{\alpha,\eta})\nonumber\\
&=&\lim_{\sigma\rightarrow 0}\Delta H_0
\partial_\Delta\sigma\,\partial_{\sigma}{\rm cov}
(\widehat{Y}_{\alpha},\widehat{Z}_{\alpha,\eta})=0.\nonumber
\end{eqnarray}

In conclusion, we see that conditions (\ref{limits}) are satisfied.


\begin{thebibliography}{35}
\bibitem{note0} To perform quantum measurements at the same
instant of time one only needs a notion of simultaneity, and not the
exact knowledge of the value of the time parameter.
\bibitem{nota} In expanding ${\cal Y}(H_0)$ around the expectation
value of $H_0$, we have implicitly admitted that the system has a
finite background energy, namely that $\langle\widehat{H}_0\rangle <
\infty$.
\bibitem{nota2} One may relax the condition of
strict positivity on $H_0^{\prime}$ and allow it to vanish, e.g., at
the single point $\pi=0$, with the function ${\cal Y}$ vanishing
hence at the corresponding pseudo energy. See in this sense the
discussion of our second set of hypotheses in Sec. IV.
\bibitem{nota3} The quantum states display
no evolution if we have adopted a Heisenberg picture (e.g., if we
have solved the quantum dynamics by introducing the appropriate
explicit time dependence in our operators).
\end{thebibliography}
\end{document}